\documentclass[11pt]{article}
\usepackage{a4wide,latexsym,graphicx,epsfig,psfrag}
\usepackage{amsmath}
\usepackage{longtable}
\usepackage{amssymb}
\usepackage{bbm}
\usepackage{cite}

\newcommand{\no}{\nonumber}

\newcommand{\lra}{\longrightarrow}

\begin{document}
\parskip=3pt plus 1pt

\begin{titlepage}
\vskip 1cm
\begin{flushright}
{\sf  UWThPh-2010-7
  \\April 2010} 
\end{flushright}

\setcounter{footnote}{0}
\renewcommand{\thefootnote}{\fnsymbol{footnote}}

\vspace*{1.5cm}
\begin{center}
{\Large\bf Chiral extrapolation and determination \\[.2cm] 
of low-energy constants from lattice data\footnote{Work
supported in part by EU contract MRTN-CT-2006-035482 (FLAVIAnet)}
} 
\\[20mm]

{\normalsize\bf G. Ecker, P. Masjuan and H. Neufeld}\\[.6cm] 
Faculty of Physics, University of Vienna \\ 
Boltzmanngasse 5, A-1090 Wien, Austria 
\end{center}

\vspace*{2cm}

\begin{abstract}
\noindent
We propose analytic approximations of chiral $SU(3)$ amplitudes for
the extrapolation of lattice data to the physical meson masses. The
method allows the determination of NNLO low-energy constants in a
controllable fashion. We test the approach with recent lattice data
for the ratio $F_K/F_\pi$ of meson decay constants.

\end{abstract}

\vfill
%
%

\setcounter{footnote}{0}
\renewcommand{\thefootnote}{\arabic{footnote}} 

\end{titlepage}


\newpage
\addtocounter{page}{1}

\paragraph{1.}
{\bf Introduction} \\[.2cm]
In recent years, lattice QCD has made enormous progress in the light
quark sector (see, e.g., Ref.~\cite{Lellouch:2009fg}). 
State-of-the-art lattice studies employ quark masses corresponding
to pion masses as low as 200 MeV, with
kaon masses close to the physical value. 

Extrapolation to the physical meson masses is performed in
different ways. On one side of the theory spectrum, there are various
smooth extrapolation formulas with more or less theoretical
motivation. On the other side, the most sophisticated extrapolations
are based on chiral 
perturbation theory (CHPT), the effective field theory of the standard
model at low energies \cite{Gasser:1983yg,Gasser:1984gg}. As the
meson masses continue to approach
the physical values in future high-statistics simulations\footnote{
  A very recent simulation \cite{Aoki:2009ix} already uses physical
  quark masses.},  
even simple-minded polynomial approximations will allow predicting
physical quantities with ever better precision. However,
a lot of information about QCD is lost in this way. On
the other hand, CHPT provides the correct analytic structure of
amplitudes in terms of several a priori undetermined constants, the
so-called low-energy constants (LECs), which are independent of the
light quark masses by definition.

Many of the higher-order LECs are difficult if not impossible to
extract from actual experimental data. Lattice calculations offer a
new environment for determining LECs because, unlike nature, the
lattice physicist can tune the quark masses. For the chiral
practitioner, it is then an advantage rather than a drawback that
present lattice studies work with different meson masses
larger than the physical values. 

Many lattice groups use next-to-leading-order (NLO) CHPT results for 
the chiral extra\-polations. As by-products, several LECs at this
order, $O(p^4)$, have actually been determined this way (see, e.g.,
Ref.~\cite{Colangelo_FLAG}). On the other hand, state-of-the-art NNLO
CHPT results have only very recently been used for the interpretation
of lattice data \cite{Bernard:2009ds,milc,Noaki:2009sk}. There are
good reasons why 
lattice physicists have generally ignored available NNLO calculations
so far:  
the results are quite involved and, what is even worse, they are
mostly available in numerical form only, at least for chiral SU(3)
(for a review of NNLO results, see Ref.~\cite{Bijnens:2006zp}). 

We propose in this note analytic approximations for NNLO CHPT
amplitudes for chiral SU(3) that are more sophisticated than the
double-log approximation \cite{Bijnens:1998yu}, yet much
simpler than the full numerical expressions. 
We first recapitulate CHPT to $O(p^6)$ for the generating functional
of Green functions \cite{Bijnens:1999hw}. The form of this functional
suggests analytic 
approximations of $p^6$-amplitudes that are scale independent in
contrast to the double-log approximation. Moreover, they include all
leading and next-to-leading contributions at large $N_c$. In a first
exploratory study, we then apply our approximation to the very
recent results of the BMW collaboration \cite{Durr:2010hr}
for the ratio $F_K/F_\pi$.

\begin{figure}[t]
\centerline{\epsfig{file=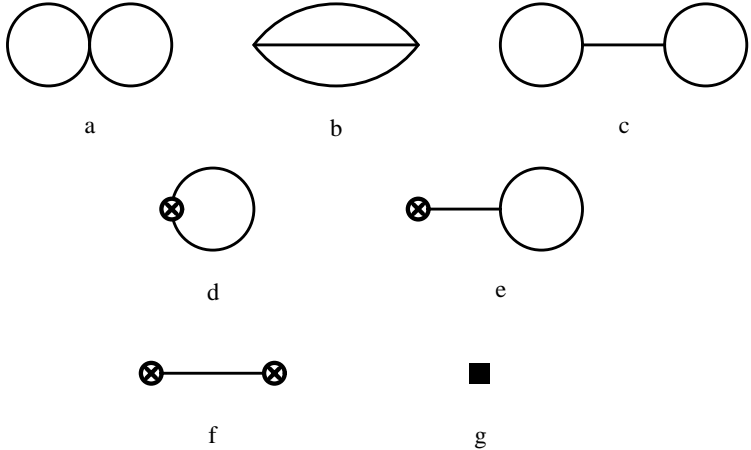,height=6cm}}
\caption{Skeleton diagrams for the generating functional
  $Z_6$ of  $O(p^6)$. Simple dots, crossed circles, black box denote
  vertices from leading-order, NLO, NNLO Lagrangians,
  respectively. Propagators and vertices carry the full tree structure 
  associated with the lowest-order Lagrangian. }
\label{fig:p6diag}
\end{figure}

\paragraph{2.}
{\bf Chiral perturbation theory to $O(p^6)$}\\[.2cm]
The most compact representation of CHPT in the meson sector is in
terms of the generating functional of Green functions $Z[j]$
\cite{Gasser:1983yg,Gasser:1984gg}. In the sequel, we suppress the
dependence on external fields $j$. Analogous to the chiral
Lagrangian, the generating functional permits a systematic chiral
expansion:
\begin{equation} 
Z = Z_2 + Z_4 + Z_6 + \dots 
\end{equation}
The NNLO functional $Z_6$ of $O(p^6)$ is itself a sum of different
contributions shown pictorially in Fig.~\ref{fig:p6diag}. In the
following, we recapitulate and reformulate the general treatment of
renormalization at $O(p^6)$ \cite{Bijnens:1999hw}.
In addition to tree diagrams of $O(p^6)$ (diagram g), there
are two classes of contributions requiring separate treatments:
irreducible (diagrams a,b,d) and reducible (diagrams c,e,f)
contributions. 

With dimensional regularization, the irreducible diagrams have both
double- and single-pole divergences. Moreover, the single-pole
divergences of each irreducible diagram are in general
non-local. Renormalization theory guarantees, however, that the sum of
the three diagrams has only local divergences
\cite{Weinberg:1978kz,Bijnens:1999hw}, i.e., polynomials in momenta
and masses 
in momentum space. Chiral symmetry guarantees that these
divergences can be absorbed by the LECs of $O(p^6)$ via diagram g. In
this process an arbitrary renormalization scale $\mu$ is
generated. The sum of diagrams a,b,d,g is then finite and can be
written in the form (details of the derivation will be given elsewhere
\cite{EMNprep})
\begin{eqnarray}
Z_6^{\rm a+b+d+g} &=&  \int \!\! d^4x \left\{ \left[C_a^r(\mu) +
\displaystyle\frac{1}{4 F_0^2} \left(4\, \Gamma_a^{(1)} \,L  -
\Gamma_a^{(2)} \,L^2  + 2\, \Gamma_a^{(L)}(\mu) L
\right)\right]  O_a(x)  \right. \no \\[.2cm] 
&&   \left.   + 
\displaystyle\frac{1}{(4\pi)^2} \left[L_i^r(\mu) - 
  \displaystyle\frac{\Gamma_i }{2} L \right] H_i(x;M) 
 +  \displaystyle\frac{1}{(4 \pi)^4}  K(x;M) \right\}~.
\label{eq:Z6irr}
\end{eqnarray} 
The structure (\ref{eq:Z6irr}) holds for chiral $SU(n)$ in general
but we have already used the notation for $n=3$. The monomials 
$O_a(x) ~(a=1,\dots,94)$ define the chiral Lagrangian of $O(p^6)$
\cite{Bijnens:1999sh} with associated renormalized LECs $C_a^r(\mu)$, 
the $L_i^r(\mu) ~(i=1,\dots,10)$ are renormalized LECs of $O(p^4)$ with
associated beta functions $\Gamma_i$ \cite{Gasser:1984gg} and the
coefficients $ \Gamma_a^{(1)}$, $\Gamma_a^{(2)}$ and $\Gamma_a^{(L)}$
are listed in Ref.~\cite{Bijnens:1999hw}. Repeated indices are to be
summed over. $F_0$ is the meson decay constant in the chiral $SU(3)$
limit. The chiral log
\begin{equation} 
L = \displaystyle\frac{1}{(4\pi)^2} \ln{M^2/\mu^2} 
\label{eq:clog}
\end{equation}
involves an additional (arbitrary) scale $M$ but $Z_6^{\rm a+b+d+g}$ as
well as the 
total generating functional are independent of both $\mu$ and
$M$. $H_i(x;M)$ are one-loop functionals associated with diagram d
whereas the two-loop contributions (except for the chiral logs) are
contained in the functional $K(x;M)$. Unlike $O_a(x)$, 
the functionals $H_i(x;M)$, $K(x;M)$ are non-local. The scale
independence of the functional (\ref{eq:Z6irr}) can be derived with
the help of renormalization group equations for the LECs
$C_a^r(\mu)$ \cite{Bijnens:1999hw} and $L_i^r(\mu)$
\cite{Gasser:1984gg}, using also relations between the coefficients
$\Gamma_a^{(2)}$ and $\Gamma_a^{(L)}$ \cite{Bijnens:1999hw}.

As shown in Ref.~\cite{Bijnens:1999hw}, the sum of reducible diagrams
c, e, f gives rise to a finite and scale independent functional
with the conventional choice of chiral Lagrangians. It can be written
in the form
\begin{eqnarray}
Z_6^{\rm c+e+f} &=&  \int \!\! d^4x\,d^4y \left[\left(L_i^r(\mu) - 
  \displaystyle\frac{\Gamma_i }{2} L \right)
  P_{i,\alpha}(x) + F_\alpha(x;M) \right]G_{\alpha,\beta}(x,y) \no
  \\[.1cm]  
&& \left[\left(L_j^r(\mu) - 
  \displaystyle\frac{\Gamma_j }{2} L \right) P_{j,\beta}(y) +
  F_\beta(y;M) \right] ~. 
\label{eq:Z6red}
\end{eqnarray}
Indices $\alpha,\beta$ run from $1,\dots,8$ (octet of
pseudoscalar mesons), the $P_{i,\alpha}(x)$ are local functionals, the
one-loop contributions of diagrams c,e are contained in the non-local
functionals $F_\alpha(x;M)$ and the $G_{\alpha,\beta}(x,y)$ are
(functional) propagators. 

The complete generating functional of $O(p^6)$ is then given by the
sum
\begin{equation}
Z_6 = Z_6^{\rm a+b+d+g} + Z_6^{\rm c+e+f}~.
\label{eq:Ztotal}
\end{equation}
Once again, it is independent of both scales $\mu$ and $M$.

\paragraph{3.}
{\bf Analytic approximation for chiral $SU(3)$}\\[.2cm]
As emphasized in the introduction, the genuine two-loop contributions
contained in the functional $K(x;M)$ are usually only available in
numerical form for chiral $SU(3)$. On the other hand, the one-loop
contributions can be given in analytic form and the dependence on
meson masses is manifest. For the chiral extrapolation of
lattice results, we therefore suggest the following approximate form
for the functional of $O(p^6)$:
\begin{eqnarray} 
Z_6^{\rm app} &=&  \int \!\! d^4x \left\{ \left[C_a^r(\mu) +
\displaystyle\frac{1}{4 F_0^2} \left(4\, \Gamma_a^{(1)} \,L  -
\Gamma_a^{(2)} \,L^2 + 2\, \Gamma_a^{(L)}(\mu) L
\right)\right] O_a(x) \right. \no \\[.2cm]
&& + \left. \displaystyle\frac{1}{(4\pi)^2} \left[L_i^r(\mu) - 
  \displaystyle\frac{\Gamma_i }{2} L \right] H_i(x;M) \right\} \no 
\\[.2cm]
&+&  \int \!\! d^4x\,d^4y \left\{\left(L_i^r(\mu) - 
  \displaystyle\frac{\Gamma_i }{2} L \right)
  P_{i,\alpha}(x) \,G_{\alpha,\beta}(x,y)  
  \left(L_j^r(\mu) - 
  \displaystyle\frac{\Gamma_j }{2} L \right) P_{j,\beta}(y)\right.
  \no \\[.1cm] 
&& + \left. 2\,\left(L_i^r(\mu) - 
  \displaystyle\frac{\Gamma_i }{2} L \right)
  P_{i,\alpha}(x) \,G_{\alpha,\beta}(x,y)\, F_\beta(y;M)\right\} ~.
\label{eq:logapp} 
\end{eqnarray}   
In contrast to the generalized double-log approximation
\cite{Bijnens:1998yu}, which can be recovered by  setting the 
coefficients $ \Gamma_a^{(1)}$ and the functionals
$H_i(x;M),\,F_\alpha(x;M)$ to zero, the analytic approximation
(\ref{eq:logapp}) is scale independent. This is an important asset for
a reliable determination of renormalized LECs.

In addition to tabulated quantities \cite{Bijnens:1999hw}, the finite
and scale 
independent one-loop functionals $H_i(x;M)$ ($i=1, \dots,10$),
$F_\alpha(x;M)$ ($\alpha=1,\dots,8$)  must be determined from diagrams
d and e. If available, the corresponding amplitudes can be read off
from existing 
calculations  by collecting all amplitudes linear in the LECs
$L_i^r(\mu)$.        

Another attractive feature of (\ref{eq:logapp}) is its
large-$N_c$ behaviour. It comprises all leading 
(generically: $C_a$, $L_i\,L_j$) and next-to-leading contributions
($L_i ~\times$ 1-loop). Of the NNLO terms, it contains at least all chiral
logs. 

The amplitude for a given observable
corresponding to the scale invariant functional (\ref{eq:logapp})
can now be determined in four steps.
\begin{enumerate}
\item Calculate all tree- and one-loop diagrams, i.e., the
  contributions from diagrams d,e,f,g in Fig.~\ref{fig:p6diag}. In
  many cases of interest, these amplitudes are already available in the
  literature \cite{Bijnens:2006zp}. 
\item In the tree-level amplitude of $O(p^6)$ (diagram g), replace the
  LECs $C_a^r(\mu)$ by
\begin{equation}
C_a^r(\mu) \lra C_a^r(\mu) + \displaystyle\frac{1}{4 F_0^2} 
\left(4\, \Gamma_a^{(1)} \,L  - \Gamma_a^{(2)} \,L^2 
+ 2\, \Gamma_a^{(L)}(\mu) L \right)~.
\label{eq:Creplace}
\end{equation}
We recall that the combination on the right-hand side of
(\ref{eq:Creplace}) is scale invariant. 
\item Collect all contributions linear and bilinear in the LECs
  $L_i^r(\mu)$ in the remaining amplitude (diagrams
  d,e,f) and extract the chiral logs. The products $L_i^r(\mu) \,L$ from
  the irreducible parts must match the terms $\Gamma_a^{(L)}(\mu) L$
  in (\ref{eq:Creplace}). After performing this check, set all chiral
  logs $L = 0$ in this subset of terms (diagrams d,e,f), which amounts
  to replacing $\mu$ by $M$ in the one-loop functions. 
\item Replace the bilinears $L_i^r(\mu) L_j^r(\mu)$ (due to reducible
  contributions: diagram f) by the scale invariant expressions 
\begin{equation}
L_i^r(\mu) L_j^r(\mu) \lra \left(L_i^r(\mu) - 
  \displaystyle\frac{\Gamma_i }{2} L \right)
\left(L_j^r(\mu) - 
  \displaystyle\frac{\Gamma_j }{2} L \right)~.
\end{equation} 
Finally, in the remaining terms linear in the $L_i^r(\mu)$
(originating from diagrams d,e) replace 
\begin{equation} 
L_i^r(\mu) \lra L_i^r(\mu) - \displaystyle\frac{\Gamma_i }{2} L~. 
\end{equation}       
\end{enumerate}
The resulting scale invariant amplitude corresponds to the functional
(\ref{eq:logapp}).  The approximation consists in dropping $K(x;M)$ and
the terms bilinear in $F_\alpha(x;M)$ in the exact functionals
(\ref{eq:Z6irr}) and (\ref{eq:Z6red}), introducing a dependence on the
scale $M$. This scale parametrizes the two-loop contributions
not contained in (\ref{eq:logapp}). Transforming the one-loop
functionals $H_i(x;M), F_\alpha(x;M)$ back to $H_i(x;\mu),
F_\alpha(x;\mu)$, the only $M$-dependence resides in the chiral logs.
The remaining (single and double) chiral logs can then
only be due to the two-loop contributions because all other
contributions are correctly included in the approximate functional
(\ref{eq:logapp}) and are therefore independent of $M$. Experience
with the double-log 
approximation \cite{Bijnens:1998yu} suggests that $M$ is naturally of
the order of the kaon mass in $SU(3)$ calculations.

While the above approximation is motivated by large $N_c$, some of the
terms not included in the approximate functional (\ref{eq:logapp})
have a relatively simple analytic form (products of one-loop amplitudes
from diagrams a,c in  Fig.~\ref{fig:p6diag}). In practice, inclusion of
those terms may improve the accuracy of the approximation for certain
observables. We will come back to this issue in Ref.~\cite{EMNprep}.

\paragraph{4.}
{\bf Application to lattice data for $F_K/F_\pi$}\\[.2cm]
We apply the analytic approximation (\ref{eq:logapp}) to the ratio
$F_K/F_\pi$ of meson decay constants. $F_K/F_\pi$ is well suited for
an exploratory study for at least two reasons.
\begin{itemize} 
\item At the scale $\mu=0.77$ GeV, the genuine two-loop contribution
  amounts to $- \,0.005$ for physical masses, i.e., half a percent only
  \cite{Amoros:1999dp,Bernard:2009ds}. 
\item The detailed results of the BMW collaboration \cite{Durr:2010hr}
  provide an ideal laboratory for testing our approximation.
\end{itemize}  
The approximate form of $F_K/F_\pi$ is given in the
Appendix where all masses are lowest-order masses of $O(p^2)$
\cite{Amoros:1999dp,HBlink}. Since we work to $O(p^6)$ the masses in
$R_4$ must be expressed in terms of the lattice masses to
$O(p^4)$ \cite{Gasser:1984gg}. The chiral limit value $F_0$ is deduced
from the experimental value $F_\pi=92.2$ MeV and physical meson
masses, using again the relation to $O(p^4)$ \cite{Gasser:1984gg}.

Here we are mainly interested in getting information on the LECs of
$O(p^6)$. Only two combinations of LECs appear:
~$C_{14} + C_{15}$ and $C_{15} + 2\,C_{17}$. Most of the $L_i$ also
contribute to $F_K/F_\pi$. Several determinations of the $L_i^r$ are
available in the literature \cite{Amoros:2001cp,Bijnens:2009hy}.
All fits of the $L_i^r$ to $O(p^6)$ 
need to make some assumptions about the $C_a^r$, in particular for
extracting $L_5^r$ from $F_K/F_\pi$. $L_5$ is the only LEC
contributing to $F_K/F_\pi$ at $O(p^4)$. Moreover, $L_5^2$ appears at
$O(p^6)$ to leading order in $1/N_c$. This suggests to 
fit the lattice data with the three parameters $L_5^r$, $C_{14}^r +
C_{15}^r$ and  $C_{15}^r + 2\,C_{17}^r$. For the remaining
LECs of $O(p^4)$ we adopt the values of fit 10 of
Ref.~\cite{Amoros:2001cp}. Since our approximation is scale
independent we may choose the conventional scale $\mu=0.77$ GeV.

Restricting the data sample of the BMW collaboration
\cite{Durr:2010hr} to simulation points with $M_\pi < 450$ MeV, we are
left with 13 data points. For this exploratory study, we take only the
statistical errors of $F_K/F_\pi$ into account. After all, our main
purpose is to investigate the capacity of lattice data for the
determination of LECs but not to compete with the
detailed analysis of Ref.~\cite{Durr:2010hr}. 

Fitting the 13 data points with Eq.~(\ref{eq:fkfpi}), we obtain for
the LECs at $\mu=0.77$ GeV
\begin{eqnarray}
L_5^r &=& (0.76 \pm 0.09)\cdot 10^{-3}      \no \\[.1cm]
C_{14}^r + C_{15}^r &=& (0.37 \pm 0.08)\cdot 10^{-3} ~{\rm GeV}^{-2}
  \no \\[.1cm]
C_{15}^r + 2\,C_{17}^r &=& (1.29 \pm 0.16)\cdot 10^{-3} ~{\rm GeV}^{-2}
~.          
\end{eqnarray}
The three parameters are strongly correlated, with correlation
coefficients indicated below.
\begin{equation} 
\begin{array}{c|cc}
   & L_5^r & C_{14}^r + C_{15}^r \\
\hline \\[-.3cm] 
C_{15}^r + 2\,C_{17}^r & 0.69 & - 0.87 \\[.2cm]
L_5^r &  & -0.95
\end{array} 
\end{equation} 
Taking these correlations into account, $F_K/F_\pi$ for physical
meson masses is found to be
\begin{equation}
F_K/F_\pi = 1.198 \pm 0.005 ~,
\end{equation}
comparing well with the result $F_K/F_\pi = 1.192(7)_{\rm
  stat}(6)_{\rm syst}$ of Ref.~\cite{Durr:2010hr}. We stress
once more that our errors take only the statistical errors of the
lattice values for $F_K/F_\pi$ into account. The
same word of caution applies to $\chi^2/{\rm dof}=1.3$ for the quality
of fit.   

Although the functional (\ref{eq:logapp}) and therefore  $F_K/F_\pi$ in
(\ref{eq:fkfpi}) are independent of the renormalization scale $\mu$
there is a residual dependence on the scale $M$ of the chiral logs.
As announced before, we adopt the natural
choice $M=M_K$ (lattice value). Varying this scale by $\pm ~20 ~\%$, both 
$F_K/F_\pi$ and $L_5^r$ remain practically unchanged while the LECs of
$O(p^6)$ vary within two standard deviations. Note that this range for
$M$ includes $M_\eta$, which is given by the Gell-Mann-Okubo mass
formula in the contribution of $O(p^6)$ and is therefore always less
than 1.2 $M_K$ for the meson masses 
under consideration. Taking $M$ too low would enhance the chiral logs
too much for an $SU(3)$ observable. The sensitivity to the scale $M$
could be substantially reduced if lattice simulations would use strange
quark masses lighter than the physical value. In such a
scenario the convergence properties of chiral $SU(3)$ could be
improved altogether. 

For the case at hand, $F_K/F_\pi$ and $L_5^r$ are insensitive to the
approximation made, with uncertainties determined by lattice
errors. The situation is opposite for the LECs of $O(p^6)$: here the
approximation errors definitely exceed the lattice errors. The
dependence on the other LECs of $O(p^4)$ must be taken into account in
addition.  

Since $C_{15}$ is subleading in $1/N_c$ our fit determines
essentially $C_{14}$ and $C_{17}$ \cite{Bernard:2009ds}. Although the
values depend of 
course on the input for the $L_i^r$ we have found generically that
both $C_{14}^r$ and $C_{17}^r$ are positive and smaller than   
$10^{-3} ~{\rm GeV}^{-2}$, always taken at the usual scale $\mu=0.77$
GeV. Comparing with resonance exchange predictions
\cite{Cirigliano:2006hb}, our results indicate that 
multi-scalar exchange is important for these LECs. With single
resonance exchange only, we would have $C_{14}^R=C_{17}^R < 0$ instead
\cite{Cirigliano:2006hb}. Our result for $L_5^r$ lies in the range
covered by other NNLO fits \cite{Amoros:2001cp,Bijnens:2009hy}.

We cannot compare directly with the results of Bernard and Passemar
\cite{Bernard:2009ds} for $C_{14}^r + C_{15}^r$ and $C_{15}^r + 2
C_{17}^r$. First of all, 
the results of the BMW collaboration were not yet available for their
analysis and, what is probably more important, the value of $L_5^r$
was taken as input in Ref.~\cite{Bernard:2009ds}. For the reasons
given earlier and because of the strong (anti-)correlations found we
consider it more appropriate to fit $L_5$
together with the LECs of $O(p^6)$. Generically, we find somewhat
bigger values for $C_{14}^r + C_{15}^r$ and $C_{15}^r + 2 C_{17}^r$
than in Ref.~\cite{Bernard:2009ds}.

\paragraph{5.} 
{\bf Conclusions} \\[.2cm]
Starting from the structure of the generating functional of Green
functions to $O(p^6)$, we have proposed analytic approximations for chiral
$SU(3)$ amplitudes that require the calculation of tree-level and
one-loop diagrams only. The result serves two purposes:
\begin{itemize} 
\item It provides flexible and user-friendly extrapolation formulas
  for lattice data. 
\item It allows for the determination of higher-order LECs that are
  otherwise difficult to extract from experimental data.
\end{itemize} 
The approximate amplitudes are independent of the
renormalization scale, a prerequisite for a reliable determination of
LECs. In addition to including all chiral logs, the amplitudes contain
all leading and next-to-leading terms in the $1/N_c$ expansion.

The approach will therefore be especially useful in cases where the
genuine two-loop contributions are small, compatible with the
large-$N_c$ counting. The ratio $F_K/F_\pi$ is an interesting
observable with this property. Fitting the approximate expression for
$F_K/F_\pi$  to recent lattice data, we obtain a value
for $F_K/F_\pi$ in agreement with the detailed analysis of
Ref.~\cite{Durr:2010hr}. Both $F_K/F_\pi$ and $L_5^r$ are insensitive
to the approximation made. The LECs of $O(p^6)$,    
$C_{14}^r + C_{15}^r$ and $C_{15}^r + 2\,C_{17}^r$, are
consistent with expectations but subject to uncertainties exceeding
the lattice errors.

Although the present study is mainly of exploratory nature we
consider the results significant enough to warrant further
investigations along these lines \cite{EMNprep}. 

\vspace*{.8cm} 

\paragraph{Acknowledgements}
We are grateful to Hans Bijnens for making the full results of
Ref.~\cite{Amoros:1999dp} accessible to us and to Laurent Lellouch for
information on the BMW data. We also thank V\'eronique Bernard, Hans
Bijnens, Gilberto Colangelo, Stefan D\"urr, Laurent Lellouch and
Emilie Passemar for helpful comments and suggestions.   
This work has been supported in part by
the EU Contract MRTN-CT-2006-035482 (FLAVIAnet).

\newpage

\begin{flushleft}  
{\bf Appendix: ~Approximate result for $F_K/F_\pi$ to $O(p^6)$}
\end{flushleft} 
In the approximation defined by the functional
(\ref{eq:logapp}), $F_K/F_\pi$ assumes the following form:
\begin{eqnarray} 
F_K/F_\pi &=& 1 + R_4 + R_6 
\label{eq:fkfpi}
\\[.3cm] 
      F_0^2 \,R_4 &=&  4 \,(M_K^2 - M_\pi^2) \,L_5 
        - 5 \,\overline{A}(M_\pi,\mu)/8
       + \overline{A}(M_K,\mu)/4 
       +  3 \,\overline{A}(M_\eta,\mu)/8 
  \no  \\[.3cm]
      F_0^4 \,R_6 &=& 8\,F_0^2 (M_K^2 - M_\pi^2) \left( 2 \,M_K^2\,
     (C_{14} + C_{15}) + M_\pi^2 \,(C_{15} + 2 \,C_{17}) \right) \no
      \\[.1cm] 
&+& (M_K^2 - M_\pi^2)\, \left( - 32 \,(M_\pi^2 + 2 \,M_K^2) L_4\,L_5  
- 8 \,(3 \,M_\pi^2 + M_K^2) L_5^2  \right. \no \\[.1cm]  
&& + \left. (25 \,M_\pi^2 + 17 \,M_K^2)L^2 /32
      \right) \no \\[.1cm]
&+& \displaystyle\frac{(M_K^2 - M_\pi^2)}{(4\pi)^2} \left( -
      2 \,(M_\pi^2 + M_K^2) L_1 -  (M_\pi^2 + M_K^2) L_2
 - (5 \,M_\pi^2 + M_K^2) L_3/18  \right. \no \\[.1cm] 
&&+\left. 6 \,(M_\pi^2 + 2 \,M_K^2) L_4 +  
(14 \,M_\pi^2 + 22 \,M_K^2) L_5/3   - 12\,(M_\pi^2 + 2 \,M_K^2) L_6  
\right. \no \\[.1cm]
&& + \left. 16 \,(M_\pi^2 - M_K^2) L_7 -  4\,(M_\pi^2 + 5 \,M_K^2) L_8 
+ (313 \,M_\pi^2 + 271 \,M_K^2) L/288
\right)  \no \\[.1cm]
&+& 5 \,\overline{A}(M_\pi,\mu)^2/8  - \overline{A}(M_K,\mu)^2/8
   + \overline{A}(M_\pi,\mu) \,\overline{A}(M_K,\mu)/16  \no \\[.1cm]    
&-& 3 \,\overline{A}(M_\pi,\mu) \,\overline{A}(M_\eta,\mu)/8 
 - 3 \,\overline{A}(M_K,\mu) \,\overline{A}(M_\eta,\mu)/16  \no \\[.1cm]
&+& \overline{A}(M_\pi,\mu) \left( 4 \,M_\pi^2 L_1 + 10 \,M_\pi^2 L_2 +
 13 \,M_\pi^2 L_3/2 + 10 \,(M_\pi^2 + 2 \,M_K^2) L_4  \right. \no
      \\[.1cm] 
&& + \left. (19 \,M_\pi^2 - 5 \,M_K^2) L_5/2 -  10 \,(M_\pi^2 + 2
      \,M_K^2) L_6 -  10 \,M_\pi^2 L_8 \right. \no \\[.1cm]
&&- \left. (361 \,M_\pi^2 + 131 \,M_K^2) L/288 \right)\no \\[.1cm]
&+& \overline{A}(M_K,\mu) \left(  - 4 \,M_K^2 L_1 - 10 \,M_K^2 L_2  
- 5 \,M_K^2 L_3 - 4 \,(M_\pi^2 + 2 \,M_K^2) L_4 
\right. \no \\[.1cm]
&&- \left. (M_\pi^2 + M_K^2) L_5 +  4\,(M_\pi^2 + 2 \,M_K^2) L_6 + 4 \,M_K^2 L_8 
+ (59 \,M_\pi^2  + 115 \,M_K^2) L/144 \right) \no \\[.1cm]
&+& \overline{A}(M_\eta,\mu) (M_K^2 - M_\pi^2)/M_\eta^2 \left(- 9 \,M_\pi^2 L_7 
- 3 \,M_\pi^2 L_8 + 5 \,M_\pi^2 L/32 \right) \no \\[.1cm]
&+& \overline{A}(M_\eta,\mu) \left( (M_\pi^2/2 - 2 \,M_K^2) L_3 
- 6\,(M_\pi^2 + 2 \,M_K^2) L_4 - (7 \,M_\pi^2 + 23 \,M_K^2) L_5/6 \right.
\no \\[.1cm] 
&&+ \left. 6 \,(M_\pi^2 + 2 \,M_K^2) L_6 + 3 \,(3 \,M_\pi^2 M_K^2/M_\eta^2 
- 7 \,M_\pi^2 + 4 \,M_K^2) L_7 \right.\no \\[.1cm] 
&&+ \left. 3 \,(M_\pi^2 M_K^2/M_\eta^2 - 3 \,M_\pi^2 + 4 \,M_K^2) L_8
      \right. \no \\[.1cm] 
&& \left. -  (15 \,M_\pi^2 M_K^2/M_\eta^2 - 44 \,M_\pi^2 - 
19 \,M_K^2) L /96  \right)
 \no
\end{eqnarray} 
The abbreviations $L_i=L_i^r(\mu)$, $C_a=C_a^r(\mu)$ have been used
for a compact representation. The masses are the lowest-order meson
masses of $O(p^2)$, $F_0$ is the meson decay constant in the chiral
$SU(3)$ limit and the chiral log $L$ is defined in (\ref{eq:clog}). The
loop function $\overline{A}(M_\alpha,\mu)$ is defined as
\begin{equation}
\overline{A}(M_\alpha,\mu)
=\displaystyle\frac{M_\alpha^2}{(4\pi)^2}
\log{\displaystyle\frac{\mu^2}{M_\alpha^2}}~.  
\end{equation}

\vspace{2cm}


\end{document}